\documentclass[10pt]{article}
\usepackage[american]{babel}
\usepackage[utf8]{inputenc}
\usepackage[T1]{fontenc}
\usepackage{times}
\usepackage{helvet}
\usepackage{courier}
\usepackage{graphicx}
\usepackage{amsmath,amssymb,amsfonts}
\usepackage{algorithm}
\usepackage{algpseudocode}
\usepackage{xcolor}
\usepackage{multirow}
\usepackage{makecell}
\usepackage{booktabs}
\usepackage{tabularx}
\newcolumntype{Y}{>{\centering\arraybackslash}X}
\newcolumntype{R}{>{\raggedleft\arraybackslash}X}
\usepackage{array}
\usepackage{tikz}
\usetikzlibrary{shapes.geometric,arrows.meta,positioning,fit,backgrounds,calc}
\usepackage{caption}
\usepackage{url}
\usepackage{seqsplit}
\usepackage{enumitem}
\usepackage{xurl}
\usepackage{microtype}
\usepackage[letterpaper,top=1in,bottom=1.125in,left=1in,right=1in,columnsep=0.3125in]{geometry}
\twocolumn

\newlength\titlebox
\makeatletter
\def\@maketitle{%
  \newpage
  \null
  \setlength{\@tempdima}{\titlebox}%
  \vskip 2em%
  \begin{center}%
    {\LARGE\bfseries \@title \par}%
    \vskip 1.5em%
    {\normalsize \@author \par}%
  \end{center}%
  \vskip 1em}
\makeatother

\usepackage{titlesec}
\titleformat{\section}{\normalfont\large\bfseries}{\thesection.}{0.5em}{}
\titleformat{\subsection}{\normalfont\normalsize\bfseries}{\thesubsection.}{0.5em}{}
\titleformat{\subsubsection}[runin]{\normalfont\normalsize\bfseries}{}{0em}{}
\titlespacing*{\section}{0pt}{7pt plus 2pt minus 2pt}{3pt}
\titlespacing*{\subsection}{0pt}{5pt plus 2pt minus 2pt}{2pt}

\usepackage{csquotes}
\usepackage[style=apa,backend=biber]{biblatex}

\newcommand{\tabnote}[1]{\par\vspace{2pt}{\footnotesize #1\par}}

\newcommand{\hicssauthor}[3]{\begin{tabular}[t]{c}{\large\textbf{#1}}\\#2\\#3\end{tabular}}
\author{%
\hicssauthor{Jeel Piyushkumar Khatiwala}{University of Baltimore}{jeel.khatiwala@ubalt.edu}\hspace{1.5em}%
\hicssauthor{Divyangkumar Patel}{Cox Automotive Inc}{divyangpatel38@gmail.com}\\[0.9em]
\hicssauthor{Dianxiang Xu}{University of Missouri-Kansas City}{dxu@umkc.edu}\hspace{1.5em}%
\hicssauthor{Weifeng Xu}{University of Baltimore}{wxu@ubalt.edu}}
\title{A Symbolic-Neural Framework for Context-Aware Evidence Extraction from Mobile SQLite Databases}
\date{}

\begin{document}
\pagestyle{empty}
\maketitle
\thispagestyle{empty}

\begin{abstract}
\textit{Reliable forensic evidence extraction from mobile SQLite databases is challenging because application schemas are undocumented, evolve across releases, and differ across operating systems, making schema-specific methods brittle and difficult to maintain. We present SNEEF, a symbolic-neural framework for context-aware evidence extraction that encodes each SQLite row as context-value pairs with standardized provenance metadata and column labels, uses a language model to recognize nine forensic entity types, and binds each accepted entity to its source table and row through the PA-EEP provenance protocol.
We evaluate SNEEF against examiner-verified ground truth from five corpora built from public Android and iOS device images, spanning multiple application versions. SNEEF achieves an F1 score of 0.967 for evidence entities extracted from SQLite row content, excluding application names, compared with 0.728 for a context-free configuration, with consistent gains across all corpora. We also introduce an Evidence Failure Mode taxonomy to characterize why extraction errors occur.}
\end{abstract}

\subsubsection*{Keywords:}
Digital forensics, SQLite databases, evidence extraction, large language models, provenance.

\section{Introduction}
\label{sec:introduction}

Smartphones are among the richest sources of digital evidence because applications continuously record communications, locations, transactions, and social activity, most often in application-specific SQLite databases~\parencite{tamma2020practical}. Extracting forensic evidence from these databases is difficult because a single seized device may contain hundreds of databases and thousands of tables, while schemas and encodings vary across applications, versions, and operating systems~\parencite{barmpatsalou2013critical}. Many application databases are also undocumented, so even the relational structure needed to interpret their contents may need to be inferred~\parencite{khatiwala2026structural}. The forensic value of an extracted artifact depends not only on recovering it but also on being able to show, later and under challenge, exactly where it came from.

Existing evidence-extraction tools face a fundamental tradeoff. Rule-based pipelines based on regular expressions and schema-specific parsers are transparent and precise while a schema remains stable, but application updates that rename columns or change timestamp encodings can break field mappings and lead to missed or misattributed artifacts~\parencite{salamh2021horizon}. Large language models (LLMs) are more flexible when processing heterogeneous content~\parencite{nayerifard2023ml,wickramasekara2025exploring}, but they can omit entities, alter extracted values, or generate unsupported content, and their outputs often lack a verifiable link to the source database row~\parencite{khatiwala2025reliability,ramadan2025jdfeg}. Existing LLM-based forensic approaches also primarily reason over documents, reports, logs, or tool-generated summaries. In contrast, SNEEF performs evidence extraction directly over SQLite rows, where contextual information and provenance can be explicitly represented and enforced.

This paper presents SNEEF, a symbolic-neural framework for context-aware evidence extraction from mobile SQLite databases. Its symbolic layer represents each SQLite row as context-value pairs that attach standardized provenance metadata (device image, database, path, table, and row line) together with the column labels of the table schema and optional per-column constraints such as declared type, timezone, or regular-expression validation. Its neural layer uses these contextualized records to recognize target forensic entity types rather than interpreting isolated values. The Provenance-Aware Evidence Extraction Protocol (PA-EEP) then binds each accepted entity to its source table and row through a deterministic identifier, enabling an examiner to verify an extracted entity through a targeted lookup rather than searching across thousands of rows.

Our contributions are threefold: (1) \textbf{Context-aware evidence extraction.} We introduce a symbolic-neural representation that encodes SQLite rows using Hierarchical Symbolic Context (HSC) and Constrained Symbolic Context (CSC), enabling evidence extraction across heterogeneous application schemas without relying on fixed field mappings; (2) \textbf{Provenance-preserving evidence extraction.} We develop the Provenance-Aware Evidence Extraction Protocol (PA-EEP), which binds each accepted entity to its source table and row for traceable verification, and use ESTRE, a recall-driven preselection stage, to reduce the number of rows requiring model-based extraction; and (3) \textbf{Failure-oriented evaluation.} We define an Evidence Failure Mode (EFM) taxonomy and Failure Mode Coverage (FMC) metric to characterize extraction failures, complementing conventional precision, recall, and F1 measures.

We evaluate SNEEF on examiner-verified ground truth from five corpora built from public Android and iOS device images, spanning multiple application versions, and release the evaluation artifacts to support reproducibility.


\section{Background and Failure Mode Definition}
\label{sec:background}

A modern forensic image routinely contains hundreds of SQLite databases and thousands of tables~\parencite{quick2014data}. Investigators must deal with undocumented schemas, encoded binary fields, fragmented and deleted records, encryption, and cross-platform variation~\parencite{barmpatsalou2013critical,barmpatsalou2018survey}. These factors strain both manual analysis and existing tools~\parencite{wu2020tools}.

Automated extraction also exhibits recurring error patterns that compromise forensic soundness. We define these patterns as Evidence Failure Modes (EFMs). Each mode is illustrated with an instance from our experiments (Section~\ref{sec:empirical_study}):

\begin{itemize}[nosep,leftmargin=1.35em,itemsep=1.5pt]
\item \textbf{FM1, Vanished Evidence:} a valid artifact is omitted. For example, album rows of a photo-library store carry a Core Data creation date (\texttt{ZCREATIONDATE}, 741026721.97, i.e., 2024-06-25 16:45:21~UTC) that the extractor leaves unreported.
\item \textbf{FM2, Unauthenticated Evidence:} a detected artifact is altered. For example, without column context the raw messaging identifier \texttt{\seqsplit{19195794674.0:0@s.whatsapp.net}}, which carries a device suffix, is reported as a phone number whose normalized digits differ from the number actually stored.
\item \textbf{FM3, Fabricated Evidence:} an artifact is generated with no support in the row. For example, the platform tag carried in the provenance metadata is echoed as an application name on rows of a sparse key-value store, a pattern documented for LLMs in forensic settings~\parencite{scanlon2023chatgpt}.
\item \textbf{FM4, Inflated Evidence:} one artifact is reported as several. For example, a contact number stored with seven dialing variants in a full-text index is emitted as seven phone numbers, one of them a seven-digit fragment.
\item \textbf{FM5, Untraceable Evidence:} a correctly extracted artifact lacks precise provenance (table, row, or offset), so the chain of custody cannot be reconstructed and the result cannot be authenticated.
\end{itemize}

Research on natural language generation distinguishes \emph{intrinsic} hallucination, which contradicts the source, from \emph{extrinsic} hallucination, which cannot be verified against the source~\parencite{ji2023hallucination,huang2025hallucination}. The EFMs map onto this distinction. Unauthenticated evidence is intrinsic. Fabricated and inflated evidence are extrinsic. Vanished evidence is the dual coverage failure, and untraceable evidence is not a hallucination but a provenance gap, which SNEEF addresses through PA-EEP rather than through extraction quality.

\section{Related Work}
\label{sec:related_work}

\textbf{Symbolic and rule-based tools.} Forensic frameworks such as the SQLite Forensic Toolkit and Autopsy rely on schema-specific parsers and regular expressions~\parencite{tamma2020practical}, and commercial artifact parsers are keyed to known application versions. \textcite{pereira2009firefox} showed that deleted browser-history records persist in unallocated space and can be carved out. \textcite{wagner2015carving} generalized record reconstruction through internal-structure carving, and later work recovers records from write-ahead logs and free lists~\parencite{meng2019bring2lite,liu2017wal,pawlaszczyk2021invisible}. These methods are effective when schemas and encodings are understood. However, they are brittle under application updates, and hand-maintained field mappings lag behind application versions~\parencite{vasilaras2022telegram}. SNEEF complements them with context-aware entity extraction under row-level traceability, evaluated under exactly the schema heterogeneity that strains rule-based parsers.

\textbf{Machine learning for forensics.} Learned models have been applied to named-entity recognition over seized textual evidence~\parencite{rodrigues2024forensicnlp} and to data reduction and mining of forensic holdings~\parencite{quick2014data}. However, surveys note that forensic machine learning has concentrated on vision and network settings rather than mobile SQLite artifacts~\parencite{nayerifard2023ml}. Such pipelines also remain evidence-type-specific and need retraining as environments shift.

\textbf{Neural and language-model reasoning.} Recent work applies LLMs to log analysis, malware analysis, and knowledge-graph construction~\parencite{wickramasekara2025exploring,akinsowon2024leveraging,khatiwala2025reliability,ramadan2025jdfeg,xu2022PST,xu2024cikm}, and early assessments documented both the promise and the hallucination risk of general-purpose models on forensic questions~\parencite{scanlon2023chatgpt}. These efforts operate over unstructured text, summaries, or tool outputs, where provenance to a concrete SQLite row is indirect. Closest to our setting, \textcite{pawlaszczyk2026text2sql} fine-tune a local model to translate an investigator's question into an executable SQLite query over forensic databases. That approach requires the investigator to know what to ask, whereas SNEEF enumerates evidence entities from every retained row and binds each to its source without a query. Adjacent work on the same data benchmarks agentic structural inference over undocumented mobile databases, whose joins can execute yet diverge from expert-verified structure~\parencite{khatiwala2026structural}, and localizes personally identifiable information across heterogeneous SQLite schemas by hypothesis-driven search~\parencite{khatiwala2026pii}. SNEEF is complementary to both: it takes the row as the unit of extraction and binds each accepted entity to its source.

\textbf{Structured-data extraction and table understanding.} Schema-as-context is well studied outside forensics. Text-to-SQL benchmarks such as Spider condition a model on a database schema to produce a query~\parencite{yu2018spider}, table-pretrained encoders such as TaPas and TaBERT jointly represent column headers and cell values~\parencite{herzig2020tapas,yin2020tabert}, and recent surveys catalogue prompting strategies for LLMs over tabular data~\parencite{fang2024tabular}. SNEEF borrows the premise that column labels disambiguate cell values, but it differs in task and output. It performs open-ended entity enumeration rather than answering a posed question, it operates row by row over thousands of heterogeneous tables, and it binds each output to a provenance identifier. The CSC constraints are guided generation~\parencite{willard2023guided} at the level of evidence types rather than tokens.

\section{The SNEEF Framework}
\label{sec:approach}

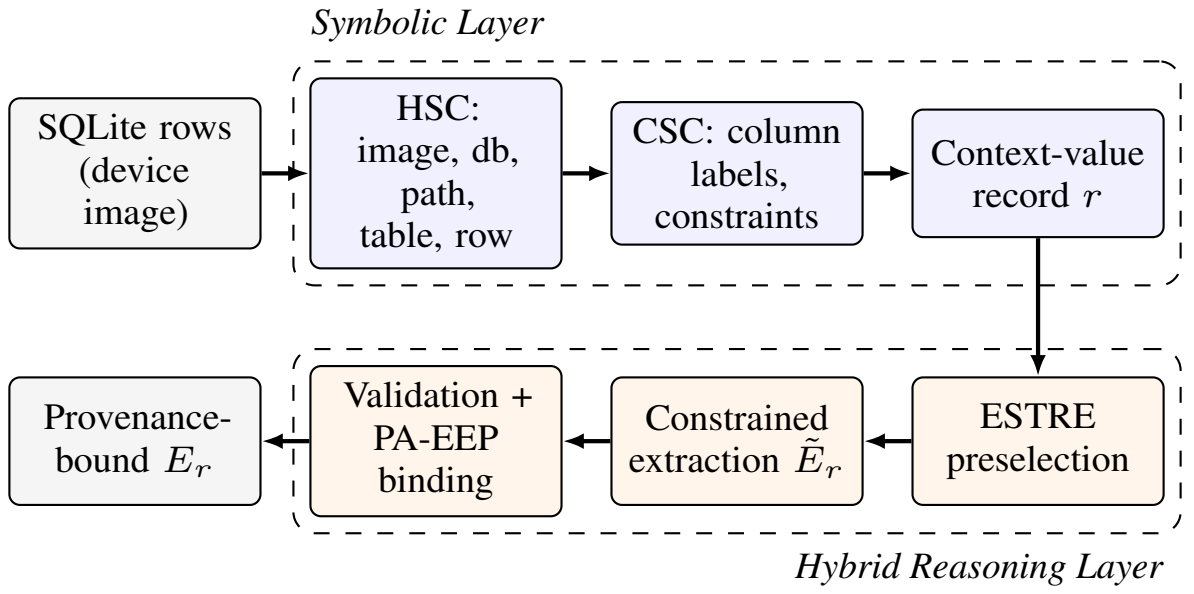
\begin{figure}[!htb]
\centering
\begin{tikzpicture}[
  font=\scriptsize,
  box/.style={draw,rounded corners=2pt,align=center,inner sep=3pt,minimum height=24pt,text width=1.45cm},
  sym/.style={box,fill=blue!6},
  neu/.style={box,fill=orange!8},
  io/.style={box,fill=black!4},
  arr/.style={-{Latex[length=4pt]},thick}
]
\node[io] (db) at (0,0) {SQLite rows\\(device image)};
\node[sym,right=9pt of db] (hsc) {HSC: image, db,\\path, table, row};
\node[sym,right=9pt of hsc] (csc) {CSC: column\\labels, constraints};
\node[sym,right=9pt of csc] (cv) {Context-value\\record $r$};
\node[neu,below=26pt of cv] (estre) {ESTRE\\preselection};
\node[neu,left=9pt of estre] (llm) {Constrained\\extraction $\tilde{E}_r$};
\node[neu,left=9pt of llm] (val) {Validation +\\PA-EEP binding};
\node[io,left=9pt of val] (out) {Provenance-\\bound $E_r$};
\draw[arr] (db) -- (hsc);
\draw[arr] (hsc) -- (csc);
\draw[arr] (csc) -- (cv);
\draw[arr] (cv) -- (estre);
\draw[arr] (estre) -- (llm);
\draw[arr] (llm) -- (val);
\draw[arr] (val) -- (out);
\begin{scope}[on background layer]
  \node[draw,dashed,rounded corners,fit=(hsc)(csc)(cv),inner sep=3pt,label={[font=\scriptsize\itshape,anchor=south west]north west:Symbolic Layer}] {};
  \node[draw,dashed,rounded corners,fit=(estre)(llm)(val),inner sep=3pt,label={[font=\scriptsize\itshape,anchor=north east]south east:Hybrid Reasoning Layer}] {};
\end{scope}
\end{tikzpicture}
\caption{The SNEEF pipeline. The Symbolic Layer builds a context-value record for each row; the Hybrid Reasoning Layer preselects evidence-bearing rows, performs constrained extraction, and binds each accepted entity to its source row through PA-EEP.}
\label{fig:overview}
\end{figure}

SNEEF comprises a \emph{Symbolic Layer} that deterministically encodes each row with standardized context and a \emph{Hybrid Reasoning Layer} that applies model-based extraction under provenance constraints (Figure~\ref{fig:overview}).

\subsection{Hierarchical Symbolic Context Modeling}
\label{sec:symbolic_context}

The Symbolic Layer converts each SQLite row into context-value pairs, so that extraction is guided by explicit metadata and labels rather than by application-specific parsing logic. Note that the encoding depends on neither column positions nor handwritten field mappings, which keeps it stable under schema drift. Each row is a set of \texttt{\{context, value\}} pairs whose contexts form a set $A = M \cup C$ with $M \cap C = \emptyset$.

The subset $M$ is the HSC, whose purpose is to state where a value comes from. It is a fixed set of five tags, $M = \{\texttt{PID}, \texttt{DN}, \texttt{DP}, \texttt{TN}, \texttt{RL}\}$: the Platform/device image Identifier, Database Name, Database Path, Table Name, and Row Line number. Together they pin each value to one image, database, table, and row; the application is read from the database path rather than carried as a separate tag. The tags are hierarchical: \texttt{PID} gives device-level context, \texttt{DN} and \texttt{DP} database-level context, \texttt{TN} table-level context, and \texttt{RL} row-level context, so the model can reason at each level of abstraction even when a schema changes between application versions.

The subset $C$ is the CSC, whose purpose is to state what a value means. It carries the schema's column labels and, optionally, per-column constraints: a declared SQLite type, a timezone or locale for temporal fields, a regular-expression validator, and a descriptive hint such as \texttt{"PHONE NUMBER"}; a \texttt{timestamp} column may be annotated as $\{\texttt{timestamp}[\textsc{integer},\textsc{est}]\}$. When constraints are unavailable, SNEEF operates on the HSC tags, the column labels, and the values alone.

A row is then the set $r = \{(a, f(a)) \mid a \in A\}$, where $f$ maps each context to its stored value. In a WhatsApp \texttt{messages} row the pair (\texttt{key\_remote\_jid}, \texttt{1650680\seqsplit{8040@s.whatsapp.net}}) tells the model that the value is a WhatsApp identifier embedding a phone number rather than free text, and (\texttt{timestamp}, \texttt{1683895200000}) identifies a millisecond epoch; a context-free prompt sees only the two strings.

\subsection{Recall-Driven Evidence Preselection}
\label{sec:selection}

ESTRE's purpose is to cut model cost without losing evidence: lightweight regular-expression detectors keep the rows likely to carry evidence and drop the rest before any model call. The thresholds follow from the cost asymmetry of the setting. An evidence-bearing row that never reaches the model is lost to the investigation, whereas an extraneous row costs only tokens and is resolved by extraction and validation. Specifically, for \emph{high-ESTRE} types whose surface form follows a pattern (phone numbers, email and IP addresses, timestamps), a detector is validated on random samples of 1\%, 5\%, and 10\% of rows and retained only if its sample recall is at least 0.90. For \emph{low-ESTRE} types without a reliable surface pattern (person names, addresses, free-form text), the target is relaxed to at least 0.50 to favor inclusion. Both thresholds are fixed before extraction and were not swept in this study (Section~\ref{sec:limitations}). For example, a high-ESTRE string such as an email address is reliably captured even when embedded in a noisy encoding such as \texttt{\seqsplit{...\textbackslash n\textbackslash x19name@gmail.com\textbackslash x12...}}, which is common in protobuf and binary fields, whereas a low-ESTRE phrase such as a street address is caught by relaxed templates that favor recall. Each retained row is truncated to a 1{,}024-token budget by dropping low-evidence fields first. Because ESTRE runs before the model, any recall measured downstream is conditional on the rows it retains.

\subsection{Provenance-Aware Symbolic-Neural Reasoning (PA-EEP)}
\label{sec:provenance}

The purpose of PA-EEP is to make every extracted entity checkable against its source: each accepted entity must carry a deterministic table identifier and a row index. The table identifier is $\mathrm{UID}(r) = \texttt{SHA256}(f(\texttt{PID}) \Vert f(\texttt{DP}) \Vert f(\texttt{TN}))$, computed with a reserved delimiter between fields so that different field splits cannot collide; with the row line index, the pair $(\mathrm{UID}(r), \texttt{RL})$ identifies exactly one row. Extraction from a record $r$ is $E_r = \texttt{PA\_EEP}(r, T)$, where $T$ is the set of target entity types and each accepted entity is
\begin{equation}
e = (\mathrm{UID}(r), l, t, v, i),
\label{eq:entity}
\end{equation}
with $l$ the row index from \texttt{RL}, $t$ the entity type, $v$ the extracted value, and $i$ a confidence score on a 1 to 10 scale that is recorded for downstream triage and not used in scoring. $T$ comprises nine forensic evidence types: application name, username, person name, phone number, email, URL keyword, message, address, and timestamp.

Algorithm~\ref{alg:paeep} summarizes PA-EEP in three stages: deterministic symbolic tagging and provenance binding, model-based entity proposal, and constraint-based validation. The neural model is confined to semantic recognition; entities that cannot be mapped to a concrete row or that violate a declared constraint are rejected. PA-EEP guarantees traceable grounding, not factual correctness beyond what the database encodes.

\begin{algorithm}[!htb]
\footnotesize
\caption{PA-EEP hybrid reasoning procedure}
\label{alg:paeep}
\begin{algorithmic}[1]
\Require Record $r$ as context-value pairs; target types $T$
\Ensure Verified evidence set $E_r$ with provenance
\Statex \textbf{Phase 1: Symbolic tagging and provenance binding}
\State Apply HSC/CSC; compute $\mathrm{UID}(r)$ as in Section~\ref{sec:provenance}
\Statex \textbf{Phase 2: Neural extraction}
\State Query the model with a constrained prompt over $r$; obtain candidates $\tilde{E}_r=\{(\tilde{t},\tilde{v},\tilde{i})\}$
\Statex \textbf{Phase 3: Validation and provenance enforcement}
\State $E_r \gets \emptyset$
\ForAll{$(\tilde{t},\tilde{v},\tilde{i}) \in \tilde{E}_r$}
  \If{$\tilde{t}\in T$ \textbf{and} $\tilde{v}$ supported by $r$ \textbf{and} constraints satisfied}
    \State $E_r \gets E_r \cup \{(\mathrm{UID}(r), \texttt{RL}, \tilde{t}, \tilde{v}, \tilde{i})\}$
  \EndIf
\EndFor
\State \Return $E_r$
\end{algorithmic}
\end{algorithm}

\section{Evaluation and Implementation Study}
\label{sec:empirical_study}

We evaluate SNEEF through four research questions. \textbf{RQ1 (Accuracy):} does context-aware encoding improve extraction accuracy across application versions and applications, and is the improvement statistically reliable? \textbf{RQ2 (Failure modes):} how does context-aware encoding change the distribution of EFMs relative to context-free extraction, counting in both directions? \textbf{RQ3 (Cost):} is the pipeline tractable at the scale of real device images? \textbf{RQ4 (Provenance):} how does PA-EEP change the cost of human verification?

\subsection{Corpora of Application Records}
\label{sec:corpora}

We built five corpora, D1 to D5, from five publicly available forensic device images. No proprietary data or commercial-tool outputs are used, so the full pipeline and ground truth can be released. Each corpus consists of the SQLite rows of the named applications, encoded as context-value records, and Table~\ref{tab:devices} lists, for each corpus, the device, the applications, the row counts before and after preprocessing, and the human-annotated ground-truth entities by type. The purpose of each corpus is as follows. D1 gives the baseline: WhatsApp's 2020 schema on Android~10. D2 tests the same application four years later, after its message table and identifiers migrated. D3 tests the same application on a different operating system. WhatsApp anchors these three corpora for three reasons. First, it is installed and populated in all three source images. Second, its stores carry the core evidence types: phone identifiers, timestamps, and message text. Third, its schema visibly migrated between the captured versions. Specifically, the Android~10 store keys messages by \texttt{key\_remote\_jid}, the Android~14 store replaces this with a \texttt{jid} table whose \texttt{user}, \texttt{server}, and \texttt{raw\_string} columns carry linked-device identifiers, and the iOS store holds Core Data offsets in \texttt{ZWAMESSAGE}. Holding the application fixed across D1 to D3 therefore isolates schema evolution across versions and operating systems. D4 tests native applications the encoding was never tuned on, and D5 tests the same synthetic persona on Android with one third-party application added. These two corpora come from one public capture-the-flag challenge. They consist of the native messaging, contacts, call-history, browser-history, and photo-library stores present in those images, plus TikTok's stores on D5, and they contribute entity types WhatsApp lacks: email addresses and download URLs. Persona and contact names in all five corpora are synthetic test data authored by the image creators.

\begin{table*}[!t]
\centering
\caption{The five corpora, each with a set of mobile applications.}
\label{tab:devices}
\footnotesize
\setlength{\tabcolsep}{3pt}
\begin{tabularx}{\textwidth}{|c|c|>{\hsize=0.85\hsize}X|c|>{\hsize=1.15\hsize}X|}
\hline
\textbf{Corpus} & \textbf{Device} & \textbf{Mobile applications} & \textbf{Rows before / after} & \textbf{Ground-truth entities} \\ \hline
D1 & Android 10 & WhatsApp: message, contact, key, and dictionary stores & 10{,}249 / 2{,}003 & 2{,}909: app 2{,}003; ts 871; phone 24; msg 11 \\ \hline
D2 & Android 14 & WhatsApp: same store families & 23{,}393 / 8{,}689 & 14{,}347: app 8{,}679; phone 3{,}615; ts 2{,}003; msg 40; person 10 \\ \hline
D3 & iOS 17 & WhatsApp: same store families, Core Data encoding & 46{,}947 / 18{,}973 & 28{,}491: app 18{,}973; ts 6{,}208; phone 2{,}403; msg 863; URL 42; person 2 \\ \hline
D4 & iOS 18 & Messages (\texttt{sms.db}), Contacts, Phone call history, Safari history, Photos & 641 / 228 & 160: ts 116; phone 23; msg 11; URL 8; email 2 \\ \hline
D5 & Android 14 & Messages (\texttt{mmssms.db}), Contacts, TikTok & 1{,}662 / 525 & 316: ts 208; URL 77; msg 26; phone 5 \\ \hline
\end{tabularx}
\tabnote{Sources: D1 and D2 from Digital Corpora~\parencite{hickman2020android10,hickman2024android14}; D3 from the Cellebrite Capture the Flag 2024 image~\parencite{cellebrite2024ctf}; D4 and D5 from the Magnet Virtual Summit 2025 CTF images~\parencite{magnet2025ctf}, both hosted by NIST CFReDS. app = application name; ts = timestamp; msg = message; URL = URL keyword; person = person name; phone = phone number.}
\end{table*}

\textbf{Preprocessing.} Each corpus is prepared in three steps. First, all rows of the named applications' databases are exported. Second, ESTRE (Section~\ref{sec:selection}) preselects the evidence-bearing rows; its detectors are validated on the 1\%, 5\%, and 10\% samples described there. Third, each retained row is truncated to the 1{,}024-token budget. As shown in Table~\ref{tab:devices}, this preprocessing reduced 82{,}892 exported rows to 30{,}418. Every retained row was encoded once with context and once without, so the protocol issued 60{,}836 model calls with no unrecovered request errors. Note that D3 alone holds 62\% of the retained rows and the three WhatsApp corpora together hold 98\%. Pooled results are therefore dominated by WhatsApp, and D4 and D5 are the smaller and harder test.

\subsection{Ground Truth and Scoring Rules}
\label{sec:groundtruth}

Ground truth covers the nine entity types of Section~\ref{sec:provenance}. A deterministic, schema-aware pass first drafted candidate entities directly from the databases. Each draft entity carries its source database, table, column, raw value, and decode rule, so every annotation is traceable to a cell. A three-examiner panel, each with five or more years of digital-forensics experience, then audited the draft in full: tracing each entity to its cell, recomputing every timestamp conversion, and adjudicating type-boundary cases against a shared codebook. Entities the draft pass had omitted but that were verbatim present in a cell were added from the database, never from any model output. Seven of the nine types are populated (Table~\ref{tab:devices}), so this is a seven-type evaluation within a nine-type schema. Address and username have no instances, because the WhatsApp stores key contacts by phone identifier and push name, and street addresses do not appear in message content.

Table~\ref{tab:scoring} states, for each entity type, the rule by which an extracted evidence entity is judged to match a ground-truth entity. This is the same judgment a human examiner would make when re-checking a row. Matching compares (type, value) pairs within a row, case-insensitively and with duplicates removed, and each type is scored on its annotated rows. The rules apply identically to both conditions. A matched entity is a true positive, an unmatched extracted entity is a false positive, and an unmatched ground-truth entity is a false negative. Timestamps in any stored form (epoch seconds or milliseconds, Core Data offsets, ISO~8601 strings) are converted to Unix UTC before comparison, so a misconverted timestamp earns no credit, and application name is treated asymmetrically because the application is a deterministic property of the database path. Two consequences follow. First, every reported recall is conditional on preselection: a ground-truth entity in a row that ESTRE removed never enters the denominator. End-to-end recall against a full-image ground truth is therefore not measured; multiplying a type's reported recall by its ESTRE acceptance threshold in Section~\ref{sec:selection} bounds it from below. Note that the contrast between conditions is unaffected, because both are scored on the same retained rows. Second, application name is deterministically derivable from the database path and is the largest annotated type. We therefore report a content-bearing aggregate that excludes it, treated as the primary measure of extraction quality, alongside the total aggregate over the seven populated types.

\begin{table}[!htb]
\centering
\caption{Ground-truth matching rules for SNEEF-discovered entities.}
\label{tab:scoring}
\footnotesize
\setlength{\tabcolsep}{4pt}
\begin{tabularx}{\columnwidth}{|p{1.8cm}|X|}
\hline
\textbf{Entity type} & \textbf{How a SNEEF-discovered entity is judged to match the ground truth} \\ \hline
Phone number & trailing ten digits equal \\ \hline
Timestamp & equal within 14~h after conversion to Unix UTC (Table~\ref{tab:tolerance} tightens this) \\ \hline
Message & same text, case-insensitive \\ \hline
URL keyword & same string, case-insensitive \\ \hline
Person name & same name, case-insensitive \\ \hline
Email & same address, case-insensitive \\ \hline
Application name & credited if reported, not missed if absent \\ \hline
\end{tabularx}
\end{table}

\subsection{Implementation}
\label{sec:implementation}

Extraction used one fixed model, Claude Opus~4.8, at temperature~0 with greedy decoding, applied identically to both conditions so that any difference is attributable to the symbolic layer. The model is named for reproducibility. Production model stacks are not bitwise-deterministic, so we describe the decoding as greedy rather than exact. In the \emph{context-aware} condition, each row carried the five HSC tags and the column labels of its table. The optional CSC constraint slots (type, timezone, regex validator, descriptive hint) were left unpopulated, so the measured gain is attributable to provenance metadata plus column labels, and the released records show the encoding exactly as the model received it. In the \emph{context-free} condition, the same rows were bare values with the row index only. Precision, recall, and F1 are pooled over rows. The 95\% confidence intervals (CIs) come from a paired percentile bootstrap with 2{,}000 resamples of rows within each corpus~\parencite{efron1993bootstrap}, and the F1 gain is the context-aware F1 minus the context-free F1 on the same rows.

\subsection{RQ1: Accuracy Across Applications and Versions}
\label{sec:rq1}

Table~\ref{tab:rq1} reports precision, recall, and F1 for each corpus; the Total row pools all five corpora over the populated entity types, and the F1 gain and its confidence interval are defined in Section~\ref{sec:implementation}. Three findings follow from the table. First, the context-aware condition improves F1 and raises recall on every corpus, and the paired bootstrap of Section~\ref{sec:implementation} places the 95\% interval of every F1 gain above zero; the widest interval, on D5, spans 0.048 to 0.173. The gain is therefore not an artifact of one corpus or of sampling. Second, the gap is largest in recall. In other words, context mainly recovers evidence that the context-free condition misses. Third, on the primary content-bearing measure of Section~\ref{sec:groundtruth}, which excludes the application-name type readable from the database path, the context-aware condition reaches F1 0.967 (95\% CI 0.965 to 0.970) against 0.728 (0.722 to 0.734), a paired difference of 0.240 (0.234 to 0.246).

\begin{table}[!htb]
\centering
\caption{Corpus-level accuracy of context-aware (CA) and context-free (CF) extraction.}
\label{tab:rq1}
\footnotesize
\setlength{\tabcolsep}{2pt}
\renewcommand{\arraystretch}{1.05}
\setlength{\tabcolsep}{2pt}\begin{tabularx}{\columnwidth}{|l|l|Y|Y|Y|Y|}
\hline
\textbf{Corpus} & \textbf{Condition} & \textbf{Precision} & \textbf{Recall} & \textbf{F1} & \makecell{\textbf{F1 gain}\\{\scriptsize(CA$-$CF)}} \\ \hline
\multirow{2}{*}{D1} & CA & 0.949 & 0.935 & \textbf{0.942} & \multirow{2}{*}{0.447} \\ \cline{2-5}
  & CF & 0.583 & 0.430 & 0.495 & \\ \hline
\multirow{2}{*}{D2} & CA & 0.976 & 0.968 & \textbf{0.972} & \multirow{2}{*}{0.178} \\ \cline{2-5}
  & CF & 0.821 & 0.769 & 0.794 & \\ \hline
\multirow{2}{*}{D3} & CA & 0.968 & 0.948 & \textbf{0.958} & \multirow{2}{*}{0.392} \\ \cline{2-5}
  & CF & 0.721 & 0.465 & 0.566 & \\ \hline
\multirow{2}{*}{D4} & CA & 0.732 & 0.631 & \textbf{0.678} & \multirow{2}{*}{0.283} \\ \cline{2-5}
  & CF & 0.741 & 0.269 & 0.394 & \\ \hline
\multirow{2}{*}{D5} & CA & 0.734 & 0.699 & \textbf{0.716} & \multirow{2}{*}{0.108} \\ \cline{2-5}
  & CF & 0.929 & 0.453 & 0.609 & \\ \hline\hline
\multirow{2}{*}{Total} & CA & 0.966 & 0.949 & \textbf{0.958} & \multirow{2}{*}{0.324} \\ \cline{2-5}
  & CF & 0.752 & 0.548 & 0.634 & \\ \hline
\end{tabularx}
\end{table}

\textbf{Observation 1: the native-application gap is mostly an annotation-representation effect.} Pooling Table~\ref{tab:rq1} by application family, the WhatsApp corpora (D1 to D3) reach F1 0.962 against 0.636, while the native-application corpora (D4, D5) reach 0.704 against 0.541, a smaller but still clear difference of 0.163 (95\% CI 0.113 to 0.216). The released outputs locate the gap. All 77 URL-keyword annotations in D5 are full asset-download URLs recorded in TikTok's \texttt{downloader.db}, and the extractor, following the schema definition of URL keyword as a search-bearing term, returned one of them. All 11 annotated messages in D4 were recorded from the \texttt{attributedBody} column of \texttt{sms.db}, a typed-stream blob whose decoded payload is the message text, whereas the extractor returned the decoded text from the parallel \texttt{text} column. The exact-match rule therefore credits none of the 11, even though the returned text is contained in each annotated blob. In D4 the extractor also left 41 of 116 timestamps unreported, 20 of them the Core Data album dates of FM1. Two sensitivity analyses quantify the annotation effects without altering the reported numbers: scoring the D4 messages as decoded text would raise D4's F1 from 0.678 to 0.752 (context-free from 0.394 to 0.459), and excluding the URL-keyword type would raise D5's from 0.716 to 0.944 (context-free from 0.609 to 0.720). The remaining shortfall is a capability limit concentrated in Core Data timestamps and SMS metadata.

\begin{table*}[!t]
\centering
\caption{Accuracy measured by entity type using all corpora, for the context-aware (CA) and context-free (CF) conditions.}
\label{tab:pertype}
\footnotesize
\setlength{\tabcolsep}{3.6pt}
\begin{tabularx}{\textwidth}{|l|R|R|Y|Y|Y|R|Y|Y|Y|}
\hline
\multirow{2}{*}{\textbf{Entity type}} & \multirow{2}{*}{\makecell{\textbf{Ground}\\\textbf{truth}}} & \multicolumn{4}{c|}{\textbf{Context-aware (CA)}} & \multicolumn{4}{c|}{\textbf{Context-free (CF)}} \\ \cline{3-10}
 & & \textbf{Discovered} & \textbf{Precision} & \textbf{Recall} & \textbf{F1} & \textbf{Discovered} & \textbf{Precision} & \textbf{Recall} & \textbf{F1} \\ \hline
Phone number & 6{,}070 & 5{,}959 & 0.999 & 0.982 & \textbf{0.991} & 5{,}179 & 0.961 & 0.853 & 0.904 \\ \hline
Timestamp & 9{,}406 & 8{,}900 & 0.987 & 0.946 & \textbf{0.966} & 4{,}009 & 0.968 & 0.426 & 0.592 \\ \hline
Application name$^{\dagger}$ & 12{,}873 / 6{,}760 & 12{,}174 & 0.945 & 0.946 & \textbf{0.945} & 3{,}087 & 0.447 & 0.457 & 0.452 \\ \hline
Message & 951 & 854 & 0.947 & 0.898 & \textbf{0.922} & 459 & 0.941 & 0.483 & 0.638 \\ \hline
URL keyword & 127 & 47 & 0.322 & 0.370 & 0.344 & 45 & 0.672 & 0.354 & \textbf{0.464} \\ \hline
Person name & 12 & 12 & \multicolumn{3}{c|}{0 spurious} & 9 & \multicolumn{3}{c|}{0 spurious} \\ \hline
Email & 2 & 2 & \multicolumn{3}{c|}{0 spurious} & 2 & \multicolumn{3}{c|}{0 spurious} \\ \hline
\end{tabularx}
\tabnote{$^{\dagger}$Scored only where the extractor reports an application name, so the ground-truth count differs by condition (12{,}873 with context, 6{,}760 without). Person name and email are given as counts only, since 12 and 2 cannot support a rate.}
\end{table*}

\textbf{Per-type results.} Table~\ref{tab:pertype} breaks the result down by entity type, pooled over the five corpora. Ground truth is the number of human-annotated entities scored, and discovered is the number of them the extractor recovered under each condition; address and username are omitted because they have no annotated instances (Section~\ref{sec:groundtruth}). The table shows where the gain comes from: the structured high-value types are extracted at high fidelity with context, and the largest single effect is on timestamp recall. Note that a timestamp is credited only when its converted value matches. The timestamp rows therefore show that the model recovers correct temporal values rather than merely noticing that a timestamp is present. URL keyword is the one type where the context-free condition scores higher. This is a definitional finding as much as a capability limit: examiners annotated whole shared and download URLs under a type defined as a search-bearing keyword, so expanded or omitted forms score as errors. In D3, the only WhatsApp corpus with the type, recall is 42 of 42 at precision 0.808; the spurious candidates are expanded forms of the same shared links, for example a \texttt{maps.app.goo.gl} link paired with its \texttt{maps.google.com/maps?q=} form.

\textbf{Timestamp tolerance.} The 14-hour window of Table~\ref{tab:scoring} was chosen to absorb timezone misinterpretation, but it crosses calendar-day boundaries. Table~\ref{tab:tolerance} therefore rescores the whole evaluation with the tolerance tightened from 14 hours down to one second. No score moves at any setting, so the generous default does not inflate the results. The extractor either converts a timestamp to within a few seconds of the stored value or misses it.

\begin{table}[!htb]
\centering
\caption{Robustness to the timestamp-matching tolerance.}
\label{tab:tolerance}
\footnotesize
\setlength{\tabcolsep}{4pt}
\renewcommand{\arraystretch}{1.0}
\begin{tabularx}{\columnwidth}{|l|Y|Y|Y|Y|}
\hline
\multirow{2}{*}{\makecell[l]{\textbf{Matching}\\\textbf{tolerance}}} & \multicolumn{2}{c|}{\makecell{\textbf{Timestamp}\\\textbf{entities (F1)}}} & \multicolumn{2}{c|}{\makecell{\textbf{Content-bearing}\\\textbf{aggregate (F1)}}} \\ \cline{2-5}
 & CA & CF & CA & CF \\ \hline
14 hours & 0.966 & 0.592 & 0.967 & 0.728 \\ \hline
1 hour & 0.966 & 0.590 & 0.967 & 0.727 \\ \hline
60 seconds & 0.966 & 0.590 & 0.967 & 0.727 \\ \hline
1 second & 0.966 & 0.589 & 0.967 & 0.726 \\ \hline
\end{tabularx}
\tabnote{CA: context-aware; CF: context-free. A match requires the absolute difference of the two Unix UTC values to be at most the tolerance. Of the 8{,}900 context-aware matches, one (a 10.8-hour offset in D3) is credited at 14 hours but not at 1 hour, six (offsets of 1.1 to 3.1 seconds in D2) are credited at 60 seconds but not at 1 second, and all others are exact to the second. The total aggregate is likewise unchanged (reproducible with the released scripts).}
\end{table}

\subsection{RQ2: Effect on Evidence Failure Modes}
\label{sec:rq2}

Accuracy summarizes outcomes. The reliability question is which \emph{kinds} of failure each configuration produces, and in which direction. A ground-truth entity is \emph{recovered by context} if it is a true positive under the context-aware condition but not under the context-free condition, and it is \emph{lost with context} in the reverse case. Both counts are reproducible from the released files. The context-aware condition recovers 15{,}736 entities the context-free condition failed to extract (5{,}049 timestamps, 842 phone numbers, 405 messages, 12 URL keywords, 3 person names, and 9{,}425 application names). In the other direction, 578 entities are credited under the context-free condition but not under the context-aware one: 338 are application names that the context-aware run did not surface at all (which the scoring rule does not penalize), and 240 are content-bearing (158 timestamps, 62 phone numbers, 10 messages, 10 URL keywords). Overall, context recovers 66 content-bearing entities for every one it loses.

The recovery is not free. Richer per-row context yields a larger pool of candidate entities before validation, and a portion of these candidates is unsupported or mistyped (the Fabricated and Inflated modes). From the run logs, returned items outside the defined schema rose from 14{,}442 in the context-free condition to 24{,}991 in the context-aware condition. These items are filtered deterministically before any entity is accepted and never reach the examiner, but their increase is the cost of the higher recall. Note that the out-of-schema counts are measured-run figures, because the released outputs are already schema-restricted; the recovery counts are reproducible.

\textbf{Observation 2: provenance metadata can itself be mistaken for content.} The platform tag ``iOS~17'' was accepted as an application name on 317 rows of D3, all but two in the near-empty \texttt{LID.sqlite} store, a Fabricated instance created by the HSC tag rather than by the row (FM3). The package-name form \texttt{com.whatsapp} on 22 D1 rows is the same error in milder form. Conversely, the context-free condition produced 207 Unauthenticated phone numbers in D2 from raw JID strings with device suffixes (FM2). None appear under context, because the column labels \texttt{user}, \texttt{server}, and \texttt{raw\_string} make the correct field explicit. In other words, context removes one class of error and introduces a smaller one, which the schema filter and validation stage contain. A corpus-wide check of email-type emissions outside the annotated rows shows the same pattern. Each condition emitted eight. With context, all eight are real addresses present in the data; without context, six of the eight are binary fragments that merely contain an @ sign.

This bidirectional, per-mode accounting is the Failure Mode Coverage of our third contribution. FMC is a reporting practice, not a score. Vanished evidence is the mode context most strongly prevents. The clearest case is D3, whose Core Data timestamps and Objective-C column names (\texttt{ZMESSAGEDATE}, \texttt{ZRECIPIENTJID}) are least self-evident; there, timestamp recall rises from 0.283 to 0.939. No untraceable evidence arises under either condition, because both are run through the PA-EEP binding.

\subsection{RQ3: Cost and Efficiency}
\label{sec:rq3}

Cost is incurred only on rows that survive preselection, so the row reductions of Table~\ref{tab:devices} translate directly into fewer model calls. From the run logs, the context-aware condition consumed 0.80M, 3.77M, 10.95M, 0.15M, and 0.32M tokens on D1 to D5, respectively, for 16.0M in total, and the complete two-condition protocol consumed 21.7M input and 4.4M output tokens across 60{,}836 calls. Mean per-record latency was stable, between 1.9 and 2.5 seconds per corpus. End-to-end time therefore scales with the retained row count rather than with schema complexity. Note that the token and latency figures are measured-run values and are not part of the released package.

\subsection{RQ4: Provenance and the Cost of Verification}
\label{sec:rq4}

Without a provenance binding, confirming a single extracted entity requires searching the originating table by hand, and this search is slowest precisely when the model has erred. PA-EEP converts the search into a targeted lookup: $\mathrm{UID}(r)$ identifies the source table, and the row index pinpoints the row, so an examiner can check the extracted value against the raw row and its neighboring metadata in one step. Every example in Sections~\ref{sec:background} and~\ref{sec:rq2} was located this way. For example, the 317 echoed platform tags resolve to \texttt{LID.sqlite} rows whose stored values contain no application name, and the 207 suffixed identifiers resolve to \texttt{jid} rows whose \texttt{user} column holds the unsuffixed number. Because the binding is present under both conditions, the failure-mode accounting of RQ2 is auditable rather than asserted. Note that we have not run a timed user study and do not claim a measured time saving.

\section{Discussion}
\label{sec:discussion}

\textbf{Why context generalizes.} A regular expression encodes a target's surface form, and it fails when that form is buried in a protobuf fragment or a platform-specific epoch. A bare-value prompt sees the value but not its role. The symbolic layer supplies the role from the schema and database path rather than from any one application's quirks, so it transfers across versions and platforms. Two extensions are left to future work: multimodal anchoring, and evidence assembled from open sources, whose effectiveness must be measured before it is relied upon~\parencite{emeksiz2026osint}.

\textbf{Operational deployment and admissibility.} The configuration evaluated here is a research proxy for benchmarking, not a deployment recommendation, because sending forensic records to an external service raises questions of jurisdiction, privacy, and chain of custody. In operational use, the pipeline would run against a locally hosted model so that evidence never leaves the controlled environment, and the HSC encoding can pseudonymize identifying fields without breaking the binding. PA-EEP gives every accepted entity a cryptographic link back to its source row, but admissibility is set by jurisdictional standards, not by a tool, so we position PA-EEP as evidence that \emph{supports} an admissibility argument. Authentication rules such as U.S. Federal Rule of Evidence 901 require showing that an item is what it is claimed to be, which a per-row binding supports, and reliability standards such as the \emph{Daubert} factors weigh a method's error rate, which the failure-mode analysis reports more directly than aggregate F1. We claim nothing about any specific jurisdiction's outcome.

\subsection{Limitations and Threats to Validity}
\label{sec:limitations}

\textbf{Baseline.} Our comparison is against a context-free ablation of the same pipeline, not against a deployed rule-based or commercial extractor. We chose this design for three reasons: commercial artifact parsers report at the artifact rather than the entity level, their outputs are proprietary, and only a same-pipeline ablation isolates the symbolic layer's contribution. The result therefore shows that symbolic context substantially improves an LLM's extraction, not that the full system outperforms a deployed tool. A public regex extractor scored under the same rules is the most valuable extension.

\textbf{What the ablation attributes.} The context-free condition removes provenance metadata and column labels together, so the measured gain is attributable to the two jointly. The optional CSC constraint slots were not populated in this evaluation (Section~\ref{sec:implementation}), so their incremental value, the framework's most forensic-specific element, is not established here. A column-labels-only condition (another 30{,}418 calls) would separate the two populated factors and is the first experiment we would add.

\textbf{Model dependence.} Results come from one hosted proprietary model, whereas Section~\ref{sec:discussion} recommends local deployment. The reported accuracy may not transfer to a local or open-weight model, no such model was evaluated, and a versioned commercial endpoint will eventually be retired, which matters for any error-rate argument under \emph{Daubert}. The released records and scorer allow both conditions to be rerun on a local model without re-annotating.

\textbf{Preselection.} Reported recall is conditional on ESTRE (Section~\ref{sec:groundtruth}). The two acceptance thresholds were fixed a priori from the cost asymmetry of missing evidence and validated on samples, not swept, and end-to-end recall against a full-image ground truth is not reported.

\textbf{Corpora and annotation.} The corpora derive from challenge-style images rather than seized casework devices, and schemas keep evolving, so generalization to live investigations is supported but not established. Three of the five corpora are one application, and the pooled results are dominated by D3. With two native-application corpora, we can speak of application coverage but not of coverage of an operating system. Seven of nine schema types are populated, and person name and email have populations of 12 and 2. Two annotation-representation choices (download URLs recorded as URL keywords, and messages recorded as typed-stream blobs) account for most of the native-application shortfall (Observation~1). The URL-keyword type is under-specified in the codebook and should be split into search keyword and shared URL.

\section{Conclusion}
\label{sec:conclusion}

We presented SNEEF, a symbolic-neural framework that extracts forensic evidence entities from mobile SQLite databases under heterogeneous schemas while binding every accepted entity to its source row, together with released evaluation artifacts from which every figure reported here can be recomputed. Representing each row as a context-value record improves LLM-based extraction over the context-free condition on every corpus tested: on content-bearing types, the aggregate F1 rises from 0.728 to 0.967, with bootstrap intervals that exclude zero and no dependence on the timestamp tolerance. The gains are conditional on preselection and are substantially a statement about WhatsApp data, and the native-application corpora are harder. Beyond accuracy, we argued that evidence extraction should be judged by how it fails, and we provided the EFM taxonomy and the bidirectional FMC account, which apply to any model-assisted evidence pipeline.

\section*{Reproducibility}

An anonymized repository at \url{https://anonymous.4open.science/r/HICSS_SNEFF-084F} holds the examiner-verified ground truth for all five corpora, the encoded records under both conditions, the schema-restricted extraction outputs, and the scoring scripts that reproduce Tables~\ref{tab:rq1}, \ref{tab:pertype}, and~\ref{tab:tolerance}, every confidence interval, and the recovery counts of Section~\ref{sec:rq2}. 

\section*{Acknowledgments}
This material is based upon work supported by the U.S. National Science Foundation under Grant Nos. 2605836 and 2556115.

\printbibliography
\end{document}